\documentclass{article}

\usepackage{arxiv}

\usepackage[utf8]{inputenc} 
\usepackage[T1]{fontenc}    
\usepackage{hyperref}       
\usepackage{url}            
\usepackage{booktabs}       
\usepackage{amsfonts}       
\usepackage{nicefrac}       
\usepackage{microtype}      
\usepackage{lipsum}
\usepackage{graphicx}

\title{Power and accountability in reinforcement learning applications to environmental policy}

\author{
  Melissa Chapman \\
  University of California Berkeley\\
  Berkeley, CA  \\
  \texttt{mchapman@berkeley.edu} \\
  \And
  Caleb Scoville \\
  Tufts University \\
  Medford, MA \\
  \texttt{caleb.scoville@tufts.edu} \\
  \And
  Marcus Lapeyrolerie \\
  University of California Berkeley\\
  Berkeley, CA \\
  \texttt{mlapeyro@berkeley.edu}  \\
   \And
  Carl Boettiger \\
  University of California Berkeley\\
  Berkeley, CA  \\
  \texttt{cboettig@berkeley.edu} \\
}

\begin{document}

\maketitle

\begin{abstract}
Machine learning (ML) methods already permeate environmental decision-making, from processing high-dimensional data on earth systems to monitoring compliance with environmental regulations. Of the ML techniques available to address pressing environmental problems (e.g., climate change, biodiversity loss), Reinforcement Learning (RL) may both hold the greatest promise and present the most pressing perils. This paper explores how RL-driven policy refracts existing power relations in the environmental domain while also creating unique challenges to ensuring equitable and accountable environmental decision processes. We leverage examples from RL applications to climate change mitigation and fisheries management to explore how RL technologies shift the distribution of power between resource users, governing bodies, and private industry. 
\end{abstract}

\section{Introduction}

An autonomous robotic submarine first splashed into the warm waters of Australia's Great Barrier Reef five years ago with a simple mission: find and destroy. Armed with 200 doses of lethal injections and capable of independent navigation and autonomous target image recognition, the COTSbot robot has become a front line of defense against the voracious Crown-of-Thorns starfish, a venomous predator which the Marine Park Authority places alongside climate change as a significant contributor to the reef's recent decline (Platt, 2016).  Artificial intelligence (AI) is already pervading the policy and practice of environmental conservation (Scoville et al., 2021).  While killer robots may raise questions about ethics and accountability in science fiction novels, the real source of such concerns is deeper beneath the surface of AI technologies and their applications. 

Machine Learning (ML) is rapidly playing a more influential role in conservation and environmental policy through a wide variety of less visible but more influential means than robotics. Importantly, ML technologies empower new stakeholders not in the visible form of anthropomorphic robots but rather the private companies (including major technology corporations) whose algorithms and infrastructures play a decisive role in the design and deployment of technology to manage and monitor common pool environmental resources. Big technology companies already maintain some of our most extensive environmental data sets (e.g., PLANET labs high-resolution satellite imagery) and increasingly develop algorithms to process those as well as publicly maintained (e.g. LANDSAT satellite imagery) data (Gorelick et al., 2017). As we explore throughout this paper, delegating control of both data and algorithms to private industry can consolidate and commercialize environmental decision-making and agenda-setting power.

Of the ML techniques available to conservation science and practice, Reinforcement Learning (RL) might hold the most promise to improve our decision-making capacity in complex environmental systems. Scientists have increasingly proposed Reinforcement learning (RL) as a method for addressing environmental management problems outside of the realm of robotics precisely because of its explicit concern with 'actions', rather than data classification and processing (Lapeyrolerie et al., 2021). RL has proven effective in finding robust strategies under dynamic and uncertain environments, precisely what many environmental policy decisions seek to accomplish. While RL-driven environmental decision-making is still primarily a proposed, rather than implemented, idea (see Table 1 and the growing library of "Conservation RL" problems at \url{boettiger-lab.github.io/conservation-gym/}), adaptive management (an iterative approach to making robust natural resource decisions under uncertainty) has long been applied in environmental management (Doremus, 2001; Doremus, 2010; Rist et al., 2013). While adaptive management ranges in its exact definition and implementation across the environmental literature, its formal definition (e.g., Chades et al., 2012) closely mirrors the structure of reinforcement learning problems, making RL-driven environmental decision making a reasonable methodology in adaptive management problems. 

But do RL-derived decisions meaningfully differ from the human-derived decisions currently used in adaptive management problems? The opacity of solutions derived from RL methods (i.e., inability to explain what aspects of the state space drive the agents' action or create counterfactual scenarios (Atrey et al., 2019)) both give power to, and reduce the accountability of, new environmental resource stakeholders: developers of RL algorithms. While human-derived solutions to adaptive natural resource management problems could present similar problems, just with different stakeholders, we suggest that RL applications in environmental decision-making also create unique considerations. The emergent issues and social implications of applying RL to domain-specific problems (Whittelstone et al., 2018), such as the trade-offs between oversight and efficacy, are particularly relevant to environmental domain applications. While a small body of existing literature has explored ethics and transparency issues in RL applications to the environmental domain (Galaz et al., 2017), this previous work has primarily responded to problems arising from the more visible applications of RL, like the COTSbot. 

This paper focuses on how reinforcement learning could reshape the terrain of power relations in environmental decision-making and policy.  Given that private industry has become integral to the operations of some of the largest-scale environmental initiatives and data sets, we explore the intersection of power, accountability, and privatization in RL-driven environmental decision processes.

\section{A brief introduction to reinforcement learning in environmental domain problems}

\begin{figure}
    \centering
    \includegraphics[width=\textwidth]{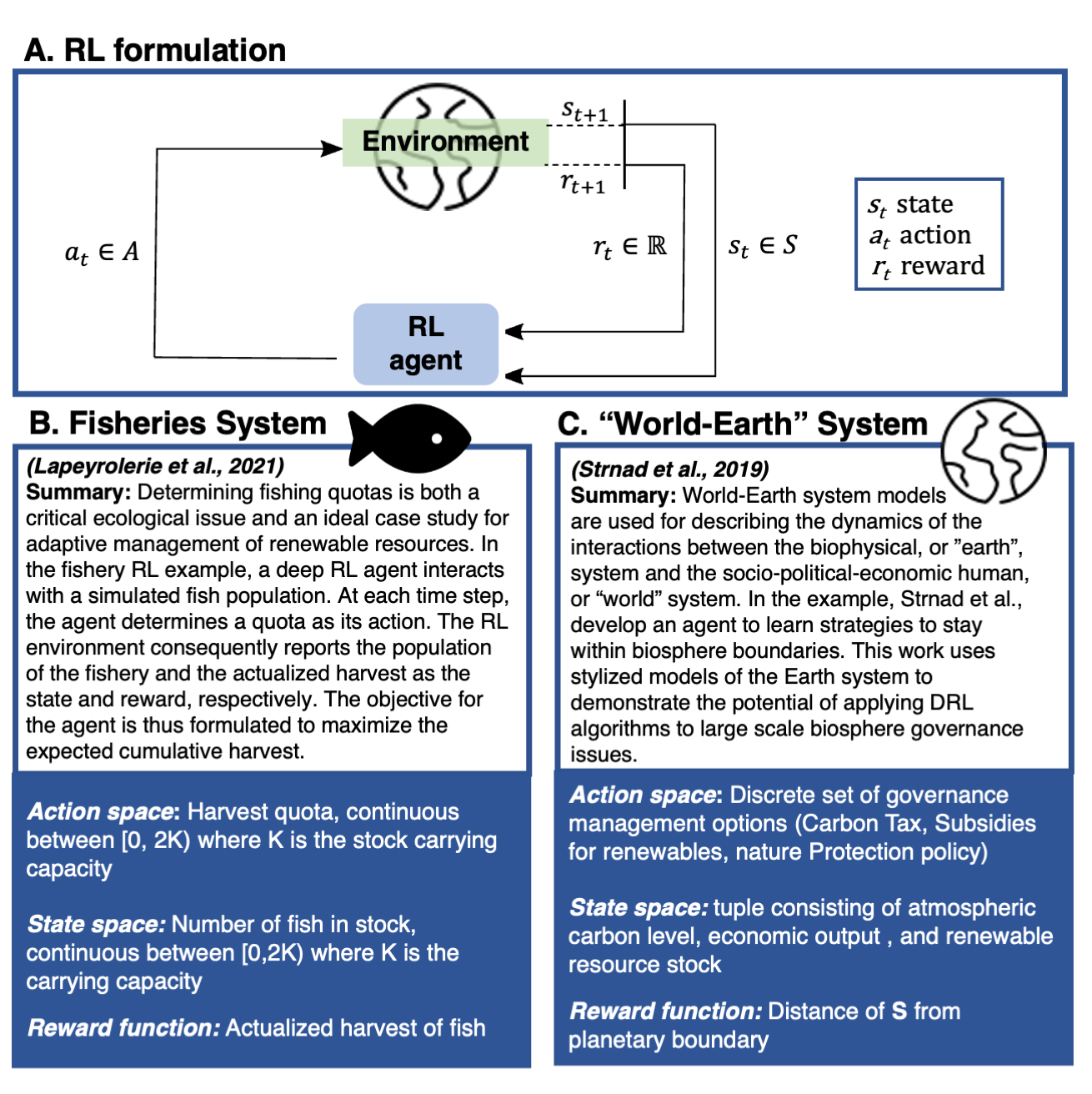}
    \caption{(A) Conceptual diagram of Reinforcement learning. RL problems consist of a set of states $S$, a set of actions $A$, set of conditional transition probabilities between states  $T(s_{t+1}|s_t, a_t)$, a reward function  $r(s_t, a_t)$, and a discount factor. The task for the agent is to learn a policy, that maps what action the agent should take given a state $s$ to maximize the expected sum of future rewards, $r$.(B-C) There are numerous proposals for applying RL to environmental problems. We highlight two examples here, which serve as examples throughout the paper. Additional examples of RL applications to environmental domain problems can be found in Table 1.}
    \label{fig:my_label}
\end{figure}

\begin{table}
    \centering
    \includegraphics[scale = 0.6]{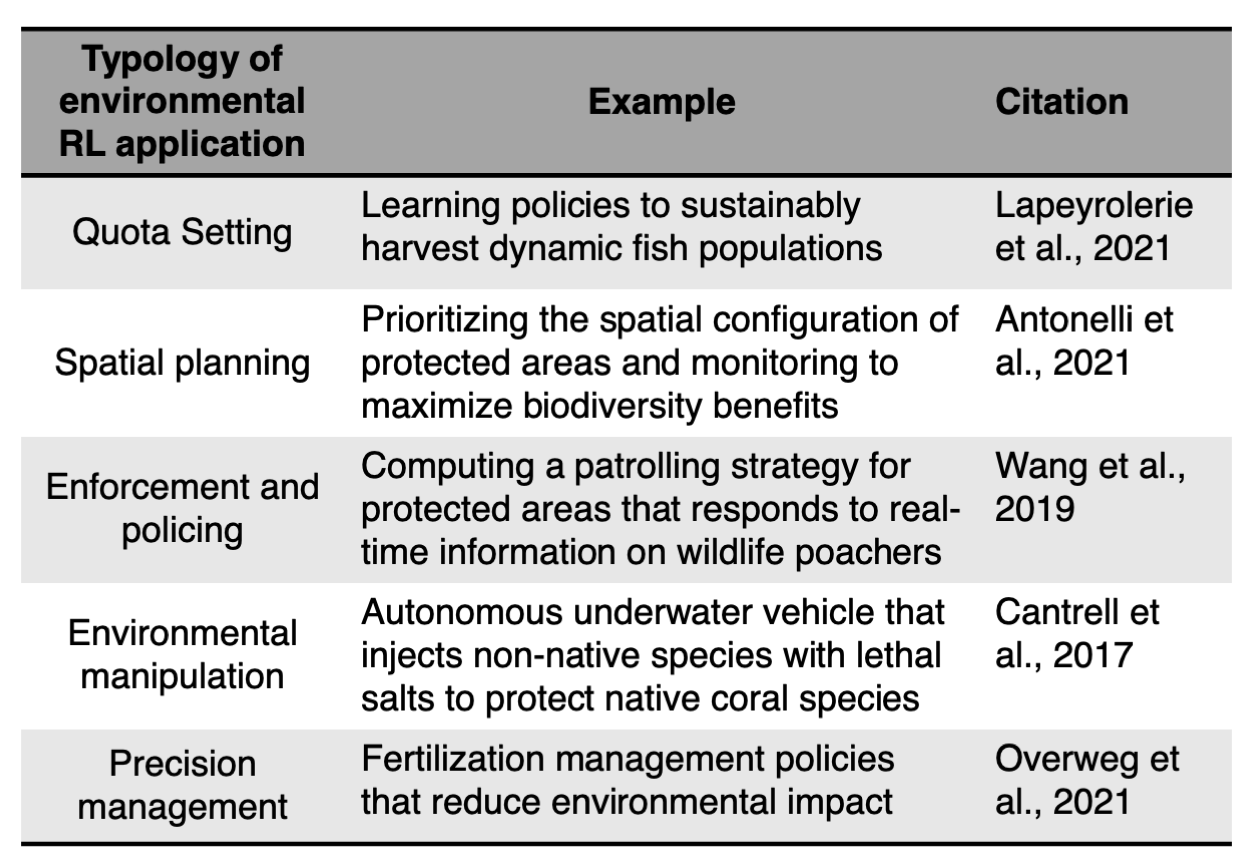}
    \caption{Examples of RL applications to environmental domain problems range from precision management of individual parcels of land to spatial planning of biodiversity monitoring and protection.}
    \label{table:my_label}
\end{table}
A thorough introduction to reinforcement learning (RL) can be found in other texts (Sutton and Barto, 2018). Here, we briefly present concepts critical to our discussion of power and accountability in environmental applications. 

Reinforcement learning is a subfield of machine learning in which an agent learns to take action in an environment to maximize some notion of reward via trial and error. When the environment is fully observable, the reinforcement learning problem can be formalized as a Markov Decision Process (MDP), which consist of a set of states $S$, a set of actions $A$, set of conditional transition probabilities between states as a function of actions, a reward function  $r(s_t, a_t)$, and a discount factor. The task for the agent is to learn a policy that maps what action the agent should take given a state $s$ to maximize the expected sum of future rewards. While there are numerous reinforcement learning algorithms to approach  finding optimal action policies, the main objective to maximize expected rewards remains (Figure 1A). It is noteworthy that deep RL, or RL that leverages deep learning methods (e.g., Neural networks) to find action policies, has excelled in solving sequential decision-making problems with complex state-action spaces, which are often intractable with classical optimization methods.

While RL is not widely applied in environmental decision making, an expanding literature suggests that there are potential applications of RL in spatial conservation planning (Antonelli et al., 2021), harvest quota setting (Lapeyrolerie et al., 2021), precision agricultural management (Overweg et al., 2021), and enforcement of environmental regulations (Wang et al.,2021) (Table 1). We focus our discussion around two proposed applications of RL in environmental decision making: (1) Harvest quota setting in fisheries and (2) global climate change mitigation strategies (Figure 1B-C).

\section{Reinforcement learning and power: Deciding on the salient features of an environmental reality}

Reinforcement learning algorithms have primarily been developed and trained in the context of games (e.g., chess) or simple physical tasks (e.g., robotics) - where distilling "reality" into a formal set of possible states and actions is comparatively uncomplicated. Most games offer a discrete set of states with defined dynamics, a set of possible actions, and rules for receiving rewards (or winning). In the case of a "game," formalizing the RL state and action space, and the reward function, is straightforward, even when learning to navigate that space effectively is not (Silver et al., 2018). Because reinforcement learning methods are still being developed predominantly in the context of these toy problems, the process of formulating environments to train and test RL algorithms has largely been uncontroversial. However, in application domains (such as environmental management), the power inherent to distilling problem scope is essential to address.

\begin{table}
    \centering
    \includegraphics[width=\textwidth]{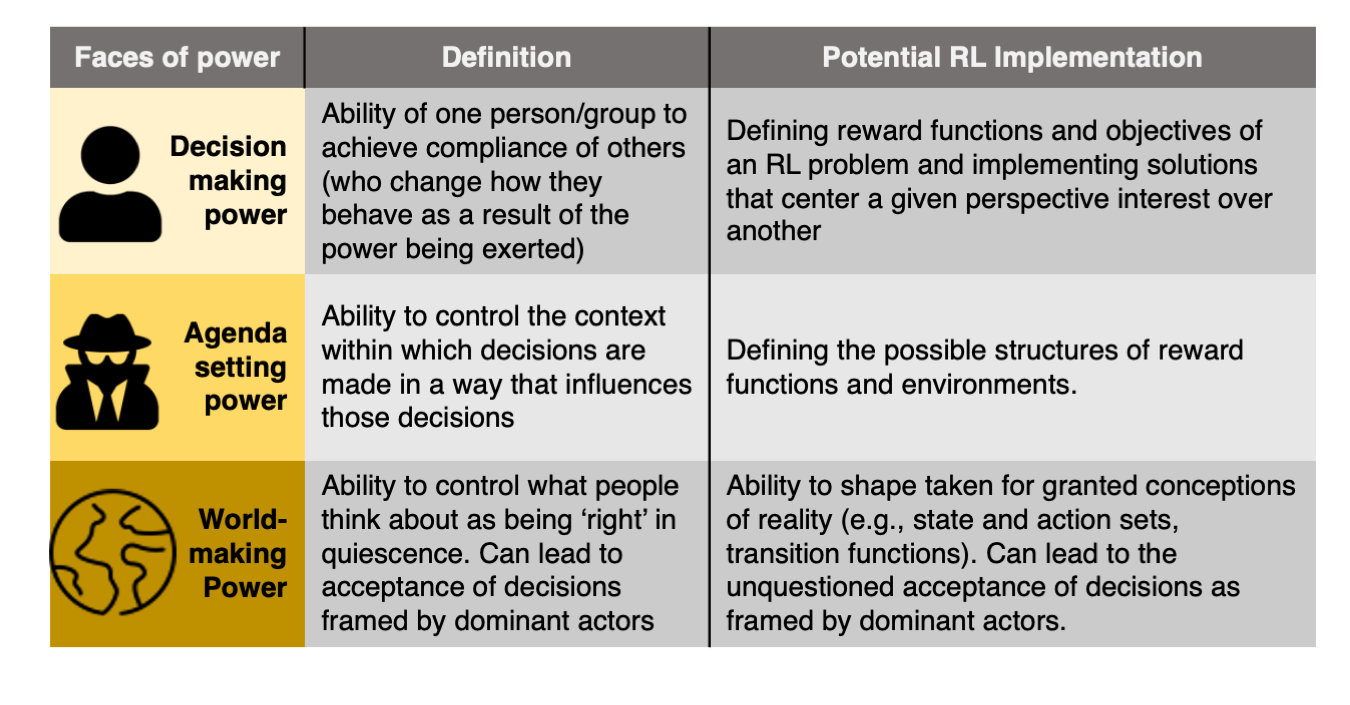}
    \caption{Three faces of power (inspired by Lukes, 2004) and their connection to RL applications to the environmental domain.}
    \label{table:my_label}
\end{table}

Applying RL to problems in the environmental domain (e.g., Figure 1B-C; Table 1) creates considerations beyond how to develop effective learning algorithms. We must first address how we define and bound that environmental problem and process. In the example of chess, the way in which we formalize the game for the agent/algorithm to learn action policies shape the "reality" that results from the agent's chosen actions. RL is "world-making" (Table 2) in the sense of conditioning "practices and capacities entailed in ordering and arranging different ways of being in the world" (Bucher, 2018:3). In domain applications, like environmental management problems, the RL agent is often first trained on some model or simulation of reality. Modeling that reality necessarily involves making assumptions about the world's salient features. However, defining the most salient features of an environmental management problem (the system dynamics, the objective, the scope of reasonable actions) is not as simple as doing so in the case of chess.  The plurality of environmental problems and processes raises the central issue herein (Levin et al., 2018). 

The world-making conditions of RL are not disconnected from the long-standing issue of representing and reconciling diverse perspectives in environmental decision-making processes. The biosphere is complex and dynamic, and as a result, reasonable differences in opinion about its reality are sure to exist among and between stakeholders and environmental decision-makers (Levin et al., 2018). The process of bounding and formalizing environmental systems effectively entrenches a particular version of what is "real" at the expense of other ways of bounding and formalizing the environment. However, unlike statistical models or scenario planning which similarly bound a complex system, RL directly implicates the decision in the conception of the world.  By defining the action space in an RL problem, we shape the scope of possibility for that environmental policy. By defining the states of the environment and the reward functions, we shape the scope of possibility for the environment the most optimal actions. When applied to a real-world setting, by shaping the state and action space and the reward process, we shape the world itself.\footnote{In this sense, RL (and potentially environmental models more generally) is "performative" in MacKenzie's (2006) sense of "effective performativity" - making a difference in the world, and potentially "Barnesian performativity" of self-validating feedback loops.}

Consider a simple example of using RL to set harvest quotas in a fishery (Figure 1B). The action space could be defined as an annual harvest quota (how many fish you are allowed to take from the sea) to maximize expected long-term yields. The state-space could be a one-dimensional representation of a given fish stock. In this case, an RL agent might be trained on a simple simulation of approximate fishery dynamics to learn an effective policy for setting quotas. After we train an agent on a simulation environment that approximates the fishery, we could query the agent's policy to find a quota for the observed stock in real life (Lapeyrolerie et al., 2021). But in an alternative representation of the problem, the state space could represent the fish stock as part of a larger ecological (or socio-ecological) system. The action space could remain the same, but the algorithm's solution to the problem in the simulation space and the real world would undoubtedly change. Alternatively, the action space could be changed (e.g. to a temporally and spatially dynamic closure of a fishery rather than a harvest quota). Like changing the state space, this would shift the environmental and social reality over time by shifting the world in which the agent is learning and making decisions. We can imagine even more options in a second example, where RL provides strategies for avoiding planetary boundaries (Figure 1C). In the case of global environmental problems, like climate change, capturing the salient features of both biophysical and socio-political-economic realities hands both agenda setting and world-making power to those implementing the methodologies even if decision making itself stays in the hands of existing governance structures (Table 2). 


In each of these examples (Figure 1B-C), it is easy to imagine where political concerns could define 'reality' (Mol, 1999): problems can be formalized to benefit the actors with the most power. But, again, these issues are not necessarily specific to or emergent from the RL technology itself. Instead, those with decision-making power already hold power to define which environmental realities are given weight and whose interests are centered. RL only refracts these issues onto a new set of stakeholders. However, RL differs from more transparent algorithmic decision making process (like an analytical model or MDP) in the relative lack of capacity to query solutions, shifting the decision making and agenda setting power to those who "define the world" in the first place.

\section{Power and Privatization}

Manipulation of preferences, behavior and political ideologies by technology (and technology companies) is widespread (Morgan, 2019). It is reasonable, then, to think that in the case of environmental decision-making processes,  private industry could exert power by not only manipulating outcomes and agendas (decision making and agenda setting power; Table 2) in favor of their interests but also by shifting the framing of realities. 

The increasingly ubiquitous private industry involvement and the potential of proprietary or partly proprietary platforms within which private industry operates create new stakeholders and shifts the balance of power in RL-driven policy. One lens for making sense of the world-making capacity of RL (and algorithmic approaches to conservation more generally) is that it may simply reshape or replace existing infrastructures of environmental policy. Private infrastructure for public policy -- or in the words of Plantin et al. (2018), the "infrastructuralization of platforms" --  poses unique problems related to stakeholder engagement, transparency and accountability. The use of RL in environmental conservation may contribute to the infrastructuralization of tech companies' platforms by making policy design and implementation reliant on particular (and potentially proprietary) technical systems. Proprietary technical systems are already used to support environmental decisions (e.g., forest and land use satellite monitoring from Planet Labs). However, unlike other technologies (e.g., ML) and data platforms, reinforcement learning centers around the formulation of decisions themselves. Figure 1C outlines a relevant example. Providing climate projections differs from the RL formulation of the problem which offers effective governance policies given some state of the system. Projections do not necessitate outlining a discrete set of potential actions. In the case of carbon taxes and subsidy allocation (Figure 1C), private industry (and tech industry specifically) has clear incentives to influence decision options to match their financial interests. 

The problem is not that tech companies might control the entire policy process through RL, or even the decision-making itself, but that their platforms mediate policy more than they previously have, and may displace other decision-making (or decision-supporting) mediums. Tech companies thus become de facto stakeholders by shaping the tools through which policies become formulated and assessed (agenda setting power; Table 2). The reliance on private companies' platforms effectively involves a delegation of the power to define policy problems away from democratically elected decision makers, even when those decision-makers still have the final say.  

\section{Accountability and blame}

The power dynamics presented by RL-driven environmental policy contribute to emergent problems in decision accountability. Even in the most advanced applications of RL (e.g., autonomous cars), the law has failed to keep up with the technology's ever-changing landscape (Greenblatt, 2016; Gless, 2016). Who is responsible for an autonomous decision that results in a devastating outcome? Who should be held accountable for a machine's actions? Referring back to our theoretical fishery example where decisions are transitioned from a simple optimal control algorithm to an RL agent, we could ask: Who is responsible if the fish stock collapses due to following decision rules suggested by an RL agent? Who is accountable for addressing when the RL action policy benefits one resource user at the cost of another, even when not immediately apparent in the RL reward function? Does accountability shift outcomes?

RL-driven decisions in the environmental domain might entrench existing problems of legally attributing accountability for environmental degradation. Cases of negligence in environmental management have a long legacy of ambiguity in the face of the law, even when considering human or corporate actors (e.g., prosecuting oil companies for the impacts of climate change). Reinforcement learning stands to distort an already ambiguous concept of "foreseeability" in environmental systems (Hunter, 2006). If companies that develop RL infrastructures for environmental decisions were to be sued for negligence (that resulted in poor or inequitable outcomes, or outcomes that served their self-interest), to place legal blame would require showing that any harm was a foreseeable consequence of negligent conduct. Because the decision policy of RL (and in particular deep RL) depends on influences external to the developer (the software evolves as it interacts with the environment and receives rewards) external forces could be pointed to as the actual cause of any bad policies or outcomes. The situation could feasibly be deemed unforeseeable and remove any liability from the RL developer. Particularly in the case of private industry management and development of RL platforms and infrastructures the issue of accountability is critical. If developers are not accountable for outcomes, policies can center self-interests over more broadly beneficial environmental results.

\section{Navigating power and accountability in RL driven environmental decisions}

As outlined above, RL-driven environmental decision-making creates novel power dynamics and important considerations for environmental accountability. We suggest the following three paths to navigate the issues raised in this piece:
\begin{enumerate}
    \item \textbf{Leverage and center participatory methods:} Inclusion of diverse voices, perspectives, values is a critical to improving the formalization of environmental problems and processes. To center the values of often-marginalized communities in the "world making" process, meaningfully consulting with communities using methods such as focus group discussions, participatory mapping, interviews, and surveys can help further include voices from resource-dependent communities (Chapman et al., 2021). Additionally, algorithmic audits by key stakeholders in a given application might help identify and prevent injustices (Raji et al., 2020). 
    \item \textbf{Maintain transparency and public engagement with environmental data and algorithms:} Avoiding reliance on proprietary software and developing open source requirements for any RL-based automated decision processes in the environmental domain is an important step towards navigating issues of power and accountability. While large technology companies have increased their control over environmental data management and algorithms throughout the past decade (e.g. Gorelick et al., 2017; Joppa, L. 2017), reclaiming public management and free and open access of data and algorithms is critical.  Because the tuning and training of RL algorithms often requires cost prohibitive amounts of computational time and power, documentation of both tuning and training processes could help ensure that objectives are not hidden underneath the nuance of RL implementation. 
    \item \textbf{Clarify channels for accountability and responsibility:} Engaging with experts in environmental law to address how RL fits into existing environmental policy statutes will be necessary to ensure responsible use of these technologies. Additionally, leveraging the broader literature on algorithms and society for lessons on developing methods for auditing RL algorithms will be critical. 
\end{enumerate}

\section{Conclusion}

RL holds promise for improving policy and management decisions in high-dimensional and uncertain environmental systems. Understanding the normative implications of building, deploying, and evaluating these technologies is critical to navigating their use in environmental policy. We show how RL applications to environmental management might refract existing power asymmetries in environmental decision-making while also creating unique problems to navigate. Finally, we raise questions about who is accountable for the outcomes of RL-driven decisions in ecological systems and how we might navigate these dimensions of RL applications to the environmental policy domain.

\section*{Acknowledgements and Disclosure of Funding}

 CB acknowledges support of the Winkler Fellowship and an NSF CAREER Award (DBI-1942280). CS acknowledges support of Tufts University CREATE Solutions (Seed Research Grant, “The Politics of Algorithmic Conservation: Artificial Intelligence and Environmental Governance in a Changing Climate").

\section*{References}

\medskip
{
\small

\begin{enumerate}
\item Adams, W. M. (2019). Geographies of conservation II: Technology, surveillance and conservation by algorithm. Progress in Human Geography, 43(2), 337-350.

\item Antonelli, A., Goria, S., Sterner, T., and Silvestro, D. (2021). Optimising biodiversity protection through artificial intelligence. bioRxiv.

\item Atrey, A., Clary, K., and Jensen, D. (2019). Exploratory not explanatory: Counterfactual analysis of saliency maps for deep reinforcement learning. arXiv preprint arXiv:1912.05743.

\item Bucher, T. (2018). If... then: Algorithmic power and politics. Oxford University Press.
    
\item Cantrell, B., Martin, L. J., and Ellis, E. C. (2017). Designing autonomy: Opportunities for new wildness in the Anthropocene. Trends in ecology and evolution, 32(3), 156-166.

\item Chades, I., Carwardine, J., Martin, T. G., Nicol, S., Sabbadin, R., and Buffet, O. (2012, July). MOMDPs: a solution for modelling adaptive management problems. In Twenty-Sixth AAAI Conference on Artificial Intelligence.

\item Chapman, M. S., Oestreich, W. K., Frawley, T. H., Boettiger, C., Diver, S., Santos, B. S., ... and Crowder, L. B. (2021). Promoting equity in the use of algorithms for high-seas conservation. One Earth, 4(6), 790-794.

\item Doremus, H. (2010). Adaptive management as an information problem. NCL Rev., 89, 1455.

\item Doremus, H. (2001). Adaptive Management, the Endangered Species Act, and the Institutional Challanges of New Age Environmental Protection. Washburn LJ, 41, 50.

\item Edwards, P. N. (2010). A vast machine: Computer models, climate data, and the politics of global warming. Mit Press.
    
\item Fang, F., Tambe, M., Dilkina, B., and Plumptre, A. J. (Eds.). (2019). Artificial intelligence and conservation. Cambridge University Press.

\item Galaz, V., Centeno, M. A., Callahan, P. W., Causevic, A., Patterson, T., Brass, I., ... and Levy, K. (2021). Artificial intelligence, systemic risks, and sustainability. Technology in Society, 67, 101741.

\item Gless, S., Silverman, E., and Weigend, T. (2016). If Robots cause harm, Who is to blame? Self-driving Cars and Criminal Liability. New Criminal Law Review, 19(3), 412-436.

\item Gorelick, N., Hancher, M., Dixon, M., Ilyushchenko, S., Thau, D., and Moore, R. (2017). Google Earth Engine: Planetary-scale geospatial analysis for everyone. Remote sensing of Environment, 202, 18-27.

\item Greenblatt, N. A. (2016). Self-driving cars and the law. IEEE spectrum, 53(2), 46-51.
    
\item Hunter, D., and Salzman, J. (2006). Negligence in the air: the duty of care in climate change litigation. U. Pa. L. Rev., 155, 1741.
    
\item Iyer, R., Li, Y., Li, H., Lewis, M., Sundar, R., and Sycara, K. (2018, December). Transparency and explanation in deep reinforcement learning neural networks. In Proceedings of the 2018 AAAI/ACM Conference on AI, Ethics, and Society (pp. 144-150).
    
\item Joppa, L. N. (2017). The case for technology investments in the environment.

\item Lapeyrolerie, M., Chapman, M. S., Norman, K. E., and Boettiger, C. (2021). Deep Reinforcement Learning for Conservation Decisions. arXiv preprint arXiv:2106.08272.

\item Lesser, L. I., Ebbeling, C. B., Goozner, M., Wypij, D., and Ludwig, D. S. (2007). Relationship between funding source and conclusion among nutrition-related scientific articles. PLoS medicine, 4(1), e5.

\item Levin, P. S., Gray, S. A., Möllmann, C., and Stier, A. C. (2021). Perception and conflict in conservation: The Rashomon effect. BioScience, 71(1), 64-72.

\item Lukes, S. (2004). Power: A radical view. Macmillan International Higher Education.

\item MacKenzie, D. (2008). An engine, not a camera: How financial models shape markets. Mit Press.

\item Mol, A. (1999). Ontological Politics. A Word and Some Questions. The Sociological Review, 74–89. doi:10.1111/j.1467-954x.1999.tb03483.x
    
\item Morgan, C. J. (2019). The Silencing Power of Algorithms: How the Facebook News Feed Algorithm Manipulates Users’ Perceptions of Opinion Climates.

\item Overweg, H., Berghuijs, H. N., and Athanasiadis, I. N. (2021). CropGym: a Reinforcement Learning Environment for Crop Management. arXiv preprint arXiv:2104.04326.

\item Plantin, J. C., Lagoze, C., Edwards, P. N., and Sandvig, C. (2018). Infrastructure studies meet platform studies in the age of Google and Facebook. New media and society, 20(1), 293-310.

\item Platt, J.R. "A Starfish-Killing, Artificially Intelligent Robot Is Set to Patrol the Great Barrier Reef". Scientific American (2016). Web. 15 Sept. 2021. 

\item Raji, I. D., Smart, A., White, R. N., Mitchell, M., Gebru, T., Hutchinson, B., ... and Barnes, P. (2020, January). Closing the AI accountability gap: Defining an end-to-end framework for internal algorithmic auditing. In Proceedings of the 2020 conference on fairness, accountability, and transparency (pp. 33-44).

\item Rist, L., Felton, A., Samuelsson, L., Sandström, C., and Rosvall, O. (2013). A new paradigm for adaptive management. Ecology and Society, 18(4).

\item Scoville, C., Chapman, M., Amironesei, R., and Boettiger, C. (2021). Algorithmic conservation in a changing climate. Current Opinion in Environmental Sustainability, 51, 30-35.

\item Silver, D., Hubert, T., Schrittwieser, J., Antonoglou, I., Lai, M., Guez, A., ... and Hassabis, D. (2018). A general reinforcement learning algorithm that masters chess, shogi, and Go through self-play. Science, 362(6419), 1140-1144.

\item Strnad, F. M., Barfuss, W., Donges, J. F., and Heitzig, J. (2019). Deep reinforcement learning in World-Earth system models to discover sustainable management strategies. Chaos: An Interdisciplinary Journal of Nonlinear Science, 29(12), 123122.

\item Sutton, R. S., and Barto, A. G. (2018). Reinforcement learning: An introduction. MIT press.

\item Wang, Y., Shi, Z. R., Yu, L., Wu, Y., Singh, R., Joppa, L., and Fang, F. (2019, July). Deep reinforcement learning for green security games with real-time information. In Proceedings of the AAAI Conference on Artificial Intelligence (Vol. 33, No. 01, pp. 1401-1408).

\item Wearn, O. R., Freeman, R., and Jacoby, D. M. (2019). Responsible AI for conservation. Nature Machine Intelligence, 1(2), 72-73.

\item Whittlestone, J., Arulkumaran, K., and Crosby, M. (2021). The Societal Implications of Deep Reinforcement Learning. Journal of Artificial Intelligence Research, 70, 1003-1030.
\end{enumerate}

}


\end{document}